\documentclass[aps,
twocolumn, 
showpacs,
pra, 
superscriptaddress,
tightenlines,
10pt] 
{revtex4-1}
\usepackage{graphicx}
\usepackage{subfig}
\usepackage{xcolor}
\usepackage{cancel}

\begin{document}
	\title{Probing the indistinguishability of single photons generated by Rydberg atomic ensembles}
	\author{Auxiliadora Padr\'on-Brito}
\affiliation{ ICFO-Institut de Ciencies Fotoniques, The Barcelona Institute of Science and Technology, 08860 Castelldefels, Barcelona, Spain}

	\author{Jan Lowinski}
\affiliation{ ICFO-Institut de Ciencies Fotoniques, The Barcelona Institute of Science and Technology, 08860 Castelldefels, Barcelona, Spain}
    \author{Pau Farrera}
\affiliation{ ICFO-Institut de Ciencies Fotoniques, The Barcelona Institute of Science and Technology, 08860 Castelldefels, Barcelona, Spain}
\affiliation{Max-Planck-Institut für Quantenoptik, Garching, Germany}

\author{Klara Theophilo}
\affiliation{ ICFO-Institut de Ciencies Fotoniques, The Barcelona Institute of Science and Technology, 08860 Castelldefels, Barcelona, Spain}
\author{Hugues de Riedmatten}
\affiliation{ ICFO-Institut de Ciencies Fotoniques, The Barcelona Institute of Science and Technology, 08860 Castelldefels, Barcelona, Spain}
\affiliation{ICREA-Instituci{\'o} Catalana de Recerca i Estudis Ava\c{c}ats, 08015 Barcelona, Spain}
%
	\date{\today}
	
\begin{abstract}
	We investigate the indistinguishability of single photons retrieved from collective Rydberg excitations in cold atomic ensembles. The Rydberg spin waves are created either by off resonant two-photon excitation to the Rydberg state or by Rydberg electromagnetically induced transparency. To assess the indistinguishability of the generated single photons, we perform Hong-Ou-Mandel experiments between the single photons and weak coherent states of light. We analyze the indistinguishability of the single photons as a function of the detection window and for photons generated by off resonance excitation we infer high value of indistinguishability going from 89$\%$ for the full waveform to 98$\%$ for small detection windows. In the same way, we also investigate for the first time the indistinguishability of single photons generated by Rydberg EIT, showing values lower than those corresponding to single photons generated by off resonance excitation. These results are relevant for the use of Rydberg atoms as quantum network nodes.
\end{abstract}

\maketitle
\section{\label{sec:intro}Introduction}

Strongly interacting Rydberg collective atomic excitations are an interesting system for realizing quantum network nodes \cite{Kimble2008}. They provide a platform allowing strong light-matter interaction without the need for a high-finesse cavity together with the capability to generate quasi-deterministic light-matter entanglement and to realize quantum processing between stored qubits \cite{Lukin2001} or between single photons \cite{Gorshkov2011,Paredes-Barato2014,Khazali2015}. Strong interaction between stored collective matter qubits is enabled by the long range interaction between Rydberg atoms \cite{Saffman2010}. Rydberg collective excitations have been used to demonstrate nonlinearities at single photon level \cite{Peyronel2012,Tiarks2016}, single photon generation using off resonant two-photon excitation to Rydberg levels \cite{Dudin2012,Ripka2018}, single photon switches and transistors \cite{Baur2014a,Gorniaczyk2014a,Tiarks2014a}, deterministic generation of ground state spin waves \cite{Li2016a,Li2016}, light-matter entanglement \cite{Li2013,Li2019}, strong interactions between distant Rydberg excitations \cite{Busche2017} and photon-photon gates with weak coherent state inputs \cite{Tiarks2019}.

For the implementation of  quantum networks using collective Rydberg excitation as quantum nodes, a crucial next step is to generate entanglement between remote nodes. For example, quantum repeater architectures have been proposed, that rely on deterministic light-matter entanglement and two-qubit gates using Rydberg ensembles 
\cite{Han2010,Zhao2010}.  
In these architectures, distant Rydberg ensembles are entangled by a measurement induced process which involves interfering single photons emitted by the Rydberg ensembles and a Bell State Measurement. For this process to be successful, the single photons must be coherent and indistinguishable, in order to erase the information about their origin before the detection. So far, only a few experiments have probed the indistinguishability of single photons emitted by Rydberg ensembles \cite{Li2013,Craddock2019,OrnelasHuerta2020}. These works assessed the photon indistinguishability by performing Hong-Ou-Mandel (HOM) experiments between the Rydberg photons and a weak coherent state \cite{Li2013}, another single photon created by a single ion \cite{Craddock2019} or another single photon created by the same Rydberg ensemble \cite{OrnelasHuerta2020}. However, in most experiments so far (with the exception of \cite{OrnelasHuerta2020}), only a small part of the temporal waveform of the Rydberg single photon was probed.  While this strategy allows observing high-visibility HOM dips for the selected photons and the generation of remote entanglement by Bell state measurements, it strongly reduces the available two photon coincidence and entanglement rate per trial, by a factor $\sim (\Delta t/T)^2$, where $\Delta t$ is the detection window considered and $T$ is the total photon duration. For these measurements, it would therefore be advantageous to use the full photon wavepacket. In addition, the works done up do date only investigated single photons generated by off resonant two photon excitation to the Rydberg state. The indistinguishability of single photons generated by Rydberg electro-magnetically induced (EIT) has not been investigated experimentally so far. 

In this paper, we report an experiment investigating the indistinguishability of photons emitted by a Rydberg atomic ensemble, both via off resonant two-photon excitation and Rydberg EIT. We perform Hong-Ou-Mandel experiments between the generated single photons and a weak coherent state and observe non-classical HOM visibilities. We analyze the indistinguishability of the single photons as a function of the detection window and infer high value of indistinguishability going from 89$\%$ (72$\%$) for the full waveform to more than 98$\%$ (87$\%$) for small detection windows, for the off resonant (EIT) case. We also discuss possible reasons for the lower indistinguishability observed for EIT photons.
		
\section{\label{sec:exp}Experimental Setup}
\begin{figure*}
	
	\centering
	\includegraphics[width=17.2cm]{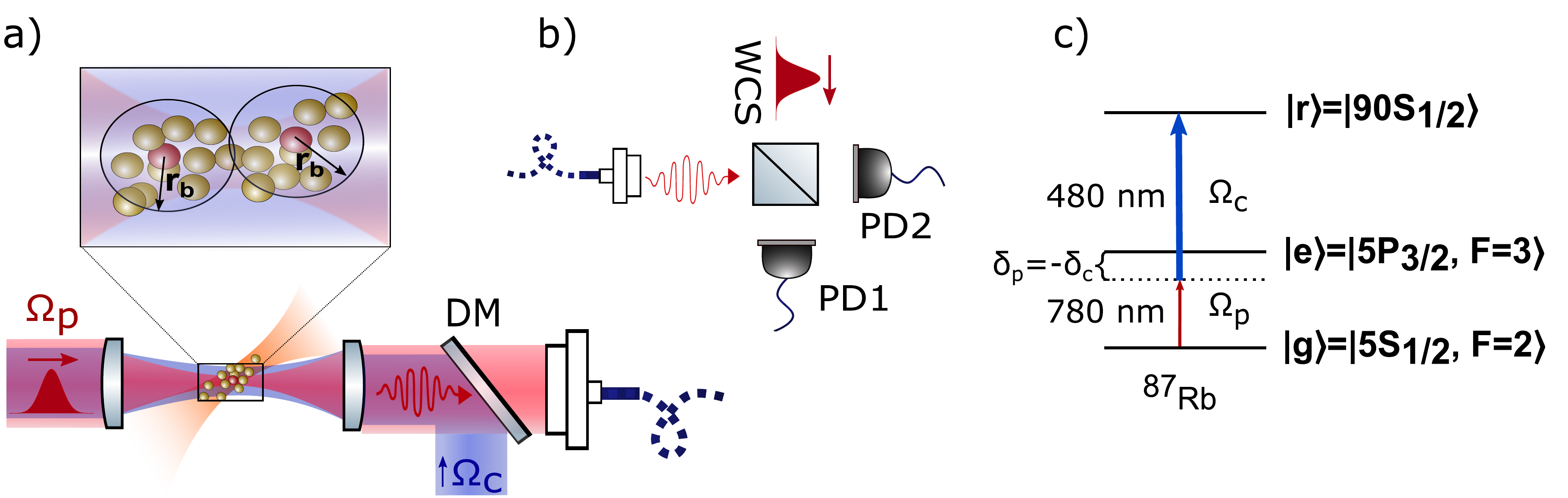}
	\captionsetup{justification=justified}
	\caption{ (a) Schematic representation of the experimental setup.  An input probe pulse (in red) is sent to the experiment with a spatial distribution given by the probe beam (also in red). The control beam (in blue) is counter-propagated with the probe and both are focused in the center of a dipole trap by two aspheric lenses. (b) The transmitted photons (curved red arrows) interfere with a WCS pulse in a HOM experiment, consisting on a beamsplitter and two single photon detectors SPD1 and SPD2. (c) Level scheme and transitions used in the experiment, where $\delta_p$ and $\delta_c$ are the detuning of the probe and control beams with regards to the intermediate level.  $\Omega_p$: Probe Rabi frequency, $\Omega_c$: Control Rabi frequency, DM: Dichroic mirror.}
	\label{fig:setup}
	
\end{figure*}

Our atomic ensemble consists of a cold cloud of $^{87}$Rb atoms. Starting from a magneto-optical trap (MOT), we compress the atoms and further cool down by optical molasses in order to efficiently load a dipole trap, which operates at $852$ nm with a transversal size of 34 $\mu$m and a depth of $\sim300 \ \mu $K, reaching a peak density of $\sim4\cdot10^{11}$ cm$^{-3}$. The experimental setup scheme and the relevant atomic levels are shown in Figure \ref{fig:setup}.

To perform the atomic excitation we turn off the dipole trap for 2.8 $\mu$s and send the excitation fields to the ensemble. The cycling of the dipole trap is repeated 37000 times at a rate of 178 kHz during the $200$ ms of the trap's lifetime. 
The atoms are initially prepared in the ground state $|g\rangle=|5S_{1/2}, F=2\rangle$, then coupled to the excited state $|e\rangle=|5P_{3/2},F'=3\rangle$ by a weak probe field at 780 nm. The excited state $|e\rangle$ is also coupled with the Rydberg state $|r\rangle=|90S_{1/2}\rangle$ by means of a counter propagating control beam at 479 nm (see Fig.\ref{fig:setup} c)). 

The one-photon detuning $\delta_c$ and $\delta_p$ are the detunings from $|e\rangle \rightarrow |r\rangle $ and $|g\rangle \rightarrow |e\rangle $ transitions, respectively, as shown in Fig.\ref{fig:setup} c). We fix the two-photon detuning $\delta_c - \delta_p\approx0$, for all experimental conditions.
The probe beam is sent with an angle of $19^o$ with respect to the dipole trap beam (see Fig.\ref{fig:setup} a)) and is focused to a beam waist of $w_p\approx6.5 \ \mu$m, in the center of the atomic medium, which has an optical depth (OD) $\sim 6$ for a resonant probe. 
  
To generate the single photons we send both probe and control fields into the medium to create a collective atomic excitation that afterwards can be retrieved as a single photon. More details of the excitation protocol are described in section \ref{sec:result}.
  
To probe the indistinguishability of the single photons emitted by the Rydberg ensemble we use the Hong-Ou-Mandel effect \cite{Hong1987} - two indistinguishable single photons impinging at a beamsplitter (BS) will bunch and exit the BS in the same output mode. In our case, we perform  HOM interference measurement between the single photons emitted by the Rydberg ensemble and weak coherent state (WCS) pulses by sending the single photon and the WCS to different input ports of the beamsplitter \cite{Rarity2005}  (see Fig.\ref{fig:setup} b). The WCS are obtained by filtering a beam derived from the same laser as the probe beam. We record the photons arrival times at each single photon detector (SPD), together with the triggers times and from that we compute the coincidences. 

To characterize the single photons, we use the second order autocorrelation function $g^{(2)}_{\Delta t}$ measured for a detection window of duration $\Delta t$, computed using the coincidences rates as:
\begin{equation}
g^{(2)}_{\Delta t}=\frac{P_c(\Delta t)}{P_1(\Delta t)P_2(\Delta t)},\label{eq:g2}
\end{equation}
where $P_c(\Delta t)$ is the coincidence detection probability and the normalization factor ${P_1(\Delta t)P_2(\Delta t)}$ is a product of the probabilities $P_{1(2)}(\Delta t)$ of a detection in detectors $1(2)$.

\section{HOM between a single photon and a weak coherent state}

To estimate the indistinguishability of the Rydberg photons from the HOM interference we follow the model presented in \cite{Li2013}. The visibility is calculated from the ratio of coincidences for distinguishable and indistinguishable inputs. To generate the distinguishable input, we apply a delay to the WCS pulse that is otherwise overlapped in time with the single photon pulse. The visibility of the HOM is defined as \begin{equation}V=1 - \frac{p_{\mathrm{ind}}}{p_{d}},\label{V}
\end{equation}
where $p_{\mathrm{ind}}$ ($p_{d}$) is the probability of a coincidence detection when the two photons are made indistinguishable (distinguishable). The probability  $p_{\mathrm{ind}}$ is computed as :
\begin{equation}
p_{\mathrm{ind}}=p_{1}^{2}g^{(2)}(0)+ \frac{1}{4}|\bar{\alpha}|^4 + (1-\eta)p_1|\bar{\alpha}|^2.\label{eqp12}
\end{equation}
Here, we consider a single photon detection probability $2p_1$ and a two-photon detection probability $p_{1}^{2}g^{(2)}(0)$, while $|\bar{\alpha}|^2$ is the mean number of photons detected on the WCS and $\eta$ is the indistinguishability factor. For completely distinguishable fields ($\eta=0$), we obtain:
\begin{equation}
p_{d}=p_{1}^{2}g^{(2)}(0)+ \frac{1}{4}|\bar{\alpha}|^4 + p_1|\bar{\alpha}|^2.
\label{eqw12}
\end{equation}
Therefore, we can express the visibility as:
\begin{equation}
V= 1- \frac{p_{\mathrm{ind}}}{p_{d}}=\frac{\eta p_1|\bar{\alpha}|^2}{p_{1}^{2}g^{(2)}(0)+ \frac{1}{4}|\bar{\alpha}|^4 + p_1|\bar{\alpha}|^2}.
\label{eqvis}
\end{equation}

From Eq.\ref{eqvis} we extract that in the limit of perfect single photon source, $g^{(2)}(0)\rightarrow0$, and low photon number in the WCS $|\bar{\alpha}|^2\rightarrow 0$, the visibility is only limited by the indistinguishability factor $\eta$. Given two completely indistinguishable photons ($\eta=1$), the third term of Eq.\ref{eqp12} describing an imperfect interference, vanishes and $\eta=1$ would lead to visibilities close to unity. 

\section{\label{sec:result}Results}
\subsection{Single photon generation}
\begin{figure} 
	\centering
	\includegraphics[width=8.5cm]{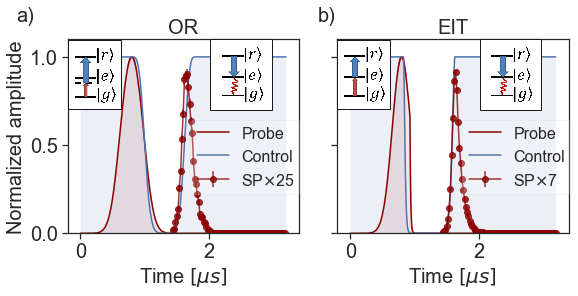}
	\captionsetup{justification=justified}
	\caption{ Representation of the probe and control optical pulses used for the photon generation under (a) OR and (b) EIT methods, normalized to their maximum value. SP shows the retrieved photon temporal distribution counts with respect to the input probe.}
	\label{fig:Pulses}
\end{figure}

First, we start by describing the pulse sequence for the single photon generation in our Rydberg ensemble. We use two different techniques, the first one is based on a two-photon off resonant (OR) excitation to the Rydberg state and the second one utilizes stopped light with EIT. In Fig. \ref{fig:Pulses} we show the temporal profiles of the input and output pulses for the OR and EIT excitation. For the OR excitation, we send a probe pulse with average photon number $n_{in}$ and a control pulse \footnote{Instead of sending a control pulse, we turn the beam off in a controllable way, as can be seen in the figure. Experimentally, sending pulses or beams in the control field does not result in significant differences in the observations.} 
into the atomic medium with both detuned from the $|e\rangle$ level by $\delta_p=-\delta_c=40$MHz . In the ideal case of an atomic cloud smaller than the blockade radius, the two-photon absorption creates a single Rydberg collective excitation, also called a Rydberg spinwave. In our case, more than one blockade sphere fits in the interaction volume, however the dephasing of multiple Rydberg excitations during the storage time ensures that only one excitation can be retrieved collectively \cite{Dudin2012}. 
This is done by sending an on-resonant control pulse after a given storage time, which trigger the emission of a single photon in the original mode of the probe. 

\begin{figure} 
	\centering
	\includegraphics[width=8.5cm]{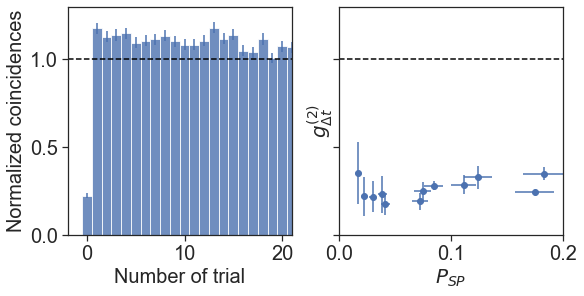}
	\captionsetup{justification=justified}
	\caption{ (a) Normalized coincidences of two different experimental trials, as a function of the number of trials between them. (b) Variation of $g^{(2)}_{\Delta t}$  as a function of  the probability of generating a single photon on the output $P_{SP}$. Both measurements were done with the OR scheme. }
	\label{fig:Generation}	
\end{figure}

The second method consists in light storage under EIT conditions. For that, we send probe (also a weak coherent state with average input photon number $n_{in}<1$)  and control pulses into the atomic medium, resonantly coupled to $|e\rangle$ level with $\delta_c=\delta_p=0$. The control pulse creates a narrow transparency window, allowing the propagation of the probe pulse throughout the cloud with a reduced group velocity $v_g$ as a dark state polariton \cite{Fleischhauer2002,Maxwell2013,Distante2016} (slow light). When the control field intensity is switched off, while the probe pulse is compressed inside the medium, $v_g$ goes to zero and the probe pulse is mapped onto a stationary Rydberg spinwave. 
After a controllable storage time in the Rydberg state, the excitation is retrieved by increasing the control field intensity to its initial value and the probe light regains its photonic characteristics and propagates (slow light) until exiting the medium and being detected.
In both cases the Rabi frequency of the control pulse is $\Omega_c=2\pi\times6 \ \mathrm{MHz}$.\\
The statistics of the generated light is analyzed with an Hanbury-Brown and Twiss setup composed of a 50/50 beamsplitter (BS) and two single photon detectors. Fig. \ref{fig:Generation}a shows the behavior of the normalized coincidences after the BS between different experimental trials, for the OR scheme. The detection window for this measurement is fixed to $\Delta t$ = 500 ns, longer than the retrieved pulse duration.
From this curve, we extract a value of $g^{(2)}_{\Delta t}$ = 0.23$\pm$0.01. The clear antibunching feature is a proof of quantum behavior and single photon emission. Slightly better antibunching is observed for EIT excitation scheme with $g^{(2)}_{\Delta t}$ = 0.17$\pm$0.01. The dependency of $g^{(2)}_{\Delta t}$ on the single photon generation probability is shown in Fig.\ref{fig:Generation}b for the OR case. The photon generation probability $P_{SP}$ is defined as $P_{SP}=\frac{p_1}{\epsilon_{det}}$ where $\epsilon_{det}$ = 0.068 is the detection efficiency which includes all optical losses from the atomic cloud to the detector including the BS for the HBT (0.157) and the SPD efficiency (0.43). $P_{SP}$ is varied by controlling the mean number of photons in the probe pulse $n_{in}$. As one can see, the source preserve its quantum characteristic with increasing number of photons with $g^{(2)}_{\Delta t} < 0.4$ for all performed measurements (with a maximal value of 0.35$\pm$0.04 for $P_{SP}$ = 0.18$\pm$0.02).  The behavior of photons generated by EIT with increasing $n_{in}$ is very different and will be discussed in section \ref{sec:EITresults}.

\subsection{Indistinguishability of photons generated by off resonant excitation.}

\begin{figure} 
	
	\centering
	\includegraphics[width=8.5cm]{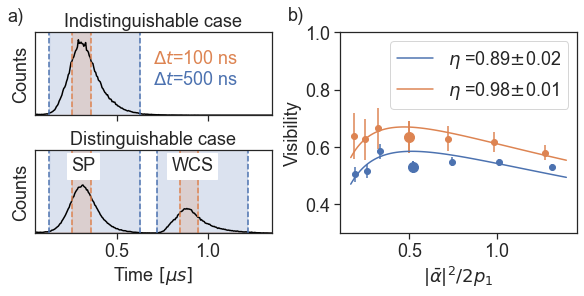}
	\captionsetup{justification=justified}
	\caption{Hong-Ou-Mandel measurements with photons generated with the OR scheme. (a) Temporal distribution of counts for the indistinguishable (top panel) and distinguishable (bottom panel) case. For the indistinguishable case, the single photon (SP) and WCS pulse overlap in time, while for the distinguishable case, a delay between them is introduced. The counts arriving in the SP window give $p_1$ and $g^{(2)}_{\Delta t}$ and the counts in the WCS window give $|\bar{\alpha}|^2$. (b) Visibility as a function of the mean number of photons in the WCS, for OR excitation, with a time observation window of $\Delta t =500 \ \mathrm{ns}$ and $\Delta t =100 \ \mathrm{ns}$.}
	\label{fig:VisOOR}
	
\end{figure}

In Fig.\ref{fig:VisOOR}b, we show the HOM visibility (calculated with Eq \ref{V}) as a function of $\frac{|\bar{\alpha}^2|}{2p_1}$ obtained by varying the mean number of photons in the WCS. The temporal distribution of the output counts is shown in Fig.\ref{fig:VisOOR}a, for both indistinguishable case (top) and distinguishable case (bottom). The values of $|\bar{\alpha}^2|$ and $p_1$ are obtained from the distinguishable measurement, by selecting the corresponding time window for the single photon (SP) and the WCS.
For the analysis of this data, we considered two different detection windows spans ($\Delta t$); one of 500 ns, which contains the full pulse and another of 100 ns, selecting only the center of the pulse. The maximum visibility achieved is 0.66$\pm$0.07 (0.58$\pm$0.03) for $\Delta t$= 100 ns ($\Delta t$= 500 ns). In both cases, we thus achieve a visibility higher than the $0.5$ classical value expected for interference of two WCS \cite{Mandel1983,Jin2013a}, attesting quantum photonic antibunching and therefore the quantum behavior of our Rydberg source.
Finally, we fit Eq. \ref{eqvis} to the data to obtain the indistinguishability factor $\eta$. For the full pulse, we find $\eta = 0.89 \pm 0.02$ and for the 100 ns window, $\eta = 0.98 \pm 0.01$. 

\begin{figure} 
	\includegraphics[width=8.5cm]{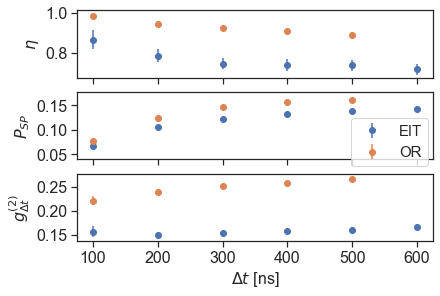}
	\caption{Variation of the indistinguishability factor $\eta$ between WCS and Rydberg photon (top panel), variation of the generation probability $P_{SP}$ of a single photon (second panel) and variation of the $g^{(2)}_{\Delta t}$ (bottom panel) as a function of the detection window $\Delta t$, for OR excitation and EIT conditions.}
	\label{fig:OORwin}
\end{figure}

It is relevant to mention that reducing the window of observation leads to higher visibilities and higher indistinguishability factors,
which are in accordance with previous observations \cite{Li2013,Craddock2019}. This is due to the fact that restricting the window of observation to values smaller than the coherence time of the retrieved photons can decrease the distinguishability between photons and therefore increase the visibility.
The analysis as a function of $\Delta t$ is shown in Fig.\ref{fig:OORwin}. On those measurements, although we observe an increase in the probability to generate a photon $P_{SP}$ with increased observation window, the decrease in the indistinguishability factor and increase of the autocorrelation function is clear, implying a statistical decrease in the quality of the single photon. This decrease could be explained by the effect of the finite laser linewidths, which limits the coherence time of both the WCS and the single photon \cite{Legero2003,Beugnon2006,Felinto2006}. Another reason could be the temporal waveform mismatch between the single photons retrieved from the Rydberg ensemble and the WCS. However, in our case we inferred a temporal indistinguishability factor above $98\%$ for OR excitation, showing that this contribution is small. Accounting for an imperfect balance of the BS \cite{Uppu2016} (47/53 in our case) results in slightly higher indistinguishability factors, with a variation much smaller than the error bars. We also did an evaluation of the accidental coincidences due to dark counts and this does not have a detrimental impact in our observations. Therefore, neither correction of BS imbalance nor subtraction of background was performed in the data presented here.

\begin{figure} 
	
	\centering
	\includegraphics[width=8.5cm]{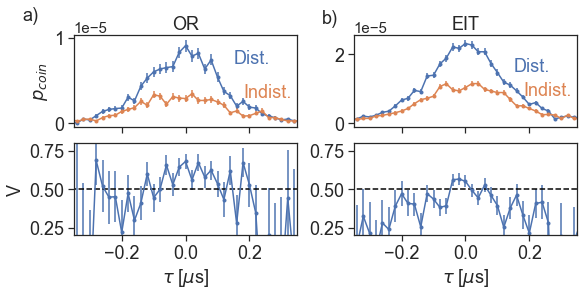}
	\caption{Time resolved coincidences for (a) OR excitation and (b) EIT excitation. On the top charts we show the time resolved coincidences, for a bin size of $20$ ns, covering the whole extension of the pulse. On the bottom charts, we show the corresponding visibilities. For those measurements $|\bar{|\alpha|}|^2/2p_1\approx0.5$ and the measured $g^{(2)}_{\Delta t}$ are 0.27$\pm$0.01 for OR and 0.17 $\pm$0.01 for EIT, respectively. }
	\label{fig:temp}
	
\end{figure}

To further evaluate how the distinguishability varies inside the photon wavepacket, we implement a time-resolved analysis, as shown in Fig \ref{fig:temp}a. We take bin sizes of $20$ ns inside the pulse and measure the  coincidences profile for the distinguishable and indistinguishable cases. We observe that the visibility is higher at the center, and decreases on the wings, which confirms the measurements of Fig. \ref{fig:OORwin}.  However, the decrease is small and the visibility, and therefore the indistinguishability, remains high over the full single photon pulse. This is in contrast to e.g. single photons emitted by single trapped ions in cavities \cite{Meraner2019} or by most solid-state systems and shows that Rydberg emitted photons are well suited for quantum communication tasks, where the possibility of using an extended detection window results in increased efficiency in the generation of entanglement between remote nodes.

\subsection{\label{sec:EITresults}Indistinguishability of EIT based single photons}

\begin{figure} 
	
	\centering
	\includegraphics[width=8.5cm]{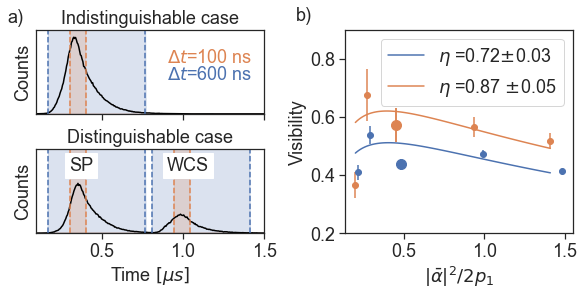}
	\captionsetup{justification=justified}
	\caption{ (a) Temporal distribution of counts for the distinguishable and indistinguishable case. For the distinguishable case a delay between single photon (SP) and WCS pulse was introduced (bottom panel), while for the indistinguishable case, both pulses overlap in time (top panel). (b) Visibility as a function the mean number of photons in the WCS, for the EIT excitation, with a time observation window of $\Delta=600$ ns (blue points)  and $\Delta=100$ ns (red points).}
	\label{fig:VisEIT}
	
\end{figure}

\begin{figure} 
 	
	\includegraphics[width=8.5cm]{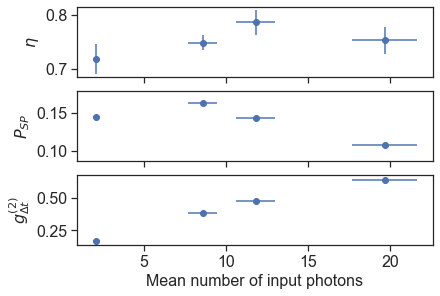}
 	\caption{  (Top panel) Variation of the indistinguishability factor $\eta$ between WCS and Rydberg photon. For that measurement, there was a frequency shift of 320 kHz between the WCS and the single photon.  (Middle panel) Variation of the generation probability of a single photon $P_{SP}$.  (Bottom panel)  Variation of the second order autocorrelation function $g^{(2)}_{\Delta t}$with the increasing in the mean number of input photons on the probe in the EIT scheme. }
 	\label{fig:EITNump}
 	
 \end{figure}

To investigate the indistinguishability of EIT photons, we performed a similar set of measurements as in the OR excitation scheme. Fig. \ref{fig:VisEIT}b shows a measurement of HOM visibility as a function of $\frac{|\bar{\alpha}^2|}{2p_1}$. Despite the fact that the value of $g^{(2)}_{\Delta t}$ is smaller for the EIT scheme, we observe that the visibility achieved is lower than for the OR case.  
It remains mostly below $0.5$ for the measurement with the whole pulse $\Delta t=600$ ns, while it reaches values slightly over $0.5$ for the smaller window $\Delta t=100$ ns. Those visibility values lead to significantly smaller indistinguishability factor $\eta$ than what we observe for OR case (see Fig. \ref{fig:VisOOR}b). Note  that for that experiment, there was a small frequency shift of 380 kHz between the WCS and the retrieved single photon. Also, the observation window for $\Delta t= 600$ ns is not centered at maximum amplitude, but optimized to include a larger portion of the pulse, because in contrast to the OR excitation, the pulse is not symmetric (see Fig. \ref{fig:VisEIT}a). The lower indistinguishability is also confirmed by the time-resolved HOM measurement  (see Fig.\ref{fig:temp}b), where we observe a similar behavior to the OR photons, but with lower visibilities. We also calculated the temporal profile mismatch between the WCS and the EIT single photons and obtained $97\%$. If frequency shifts between WCS and single photon central frequency are considered, this value is decreased to $94\%$. These values, while slightly lower than for the OR case, are not sufficient to explain the lower indistinguishability of EIT generated photons.\\
The disrupted photon generation may be explained considering that, in Rydberg EIT conditions, input photons can interact with the intermediate excited state and lead to photon scattering depending on the input photon number $n_{in}$.
For the light transmitted under the regime of Rydberg EIT, the photon purity $P$ (which represents an upper bound for the indistinguishability $\eta$) due to multiphoton scattering  has been predicted theoretically to  decrease with $n_{in}$,  starting from 1 for $n_{in} \ll 1$ and tending to 0.5 for $n_{in}>$ 5 \cite{Gorshkov2013}. 
This loss of photon purity is due to the fact that when $n_{in}$ increases, there is an increasing probability of finding two photons in the input pulse separated by less than a blockade radius. In the ideal case, the first photon is mapped onto an Rydberg polariton, which blocks the transmission of subsequent photons within a blockade radius which causes photon scattering. The leakage of information into the environment will then project the Rydberg polariton onto a mixture of localized spinwaves shorter than the blockade time, defined as $\tau_b=r_b/v_{g}$. To  investigate the impact of the multiphoton scattering inside the probe pulse and pollutants creation, we performed the HOM measurements with varying number of input photons in the probe pulse. The results are shown in Fig.\ref{fig:EITNump}. We observe that the indistinguishability $\eta$ varies between 0.7 and 0.8 for  $n_{in}$ varying between 1 and 20.  While this seems in contradiction with the model presented in \cite{Gorshkov2013}, we note that this model is designed for traveling slow light polaritons while in our case we use storage in Rydberg states. The discrepancy might also be explained by the fact that ref \cite{Gorshkov2013} did not take into account the effect of filtering due to the EIT transparency window, which was later studied in \cite{Zeuthen2017}. When  $n_{in}$ increases, the distance between two photons in the input pulse decreases. Therefore the photon scattering will make the first polariton wavefunction more localized in time, with a frequency spectrum that can potentially exceed the  EIT transparency linewidth. This effect leads to a frequency filtering of the emitted photons increasing with $n_{in}$,  which could increase  the indistinguishability of the transmitted photons and  explain why $\eta$ does not drop with $n_{in}$.  In addition to the indistinguishability, we also measured  the probability to generate a single photon $P_{SP}$ and the  $g^{(2)}_{\Delta t}$ as a function $n_{in}$, as shown in Fig.\ref{fig:EITNump} b and c. We observe, that $P_{SP}$ decreases with increasing number of input photons, particularly for $n_{in}>10$.  This behavior is qualitatively consistent with the EIT filtering effect mentioned above. 

The decrease in efficiency is accompanied by a strong degradation of the single photon quality,  witnessed by a large increase in $g^{(2)}_{\Delta t}$, up to $g^{(2)}_{\Delta t}$ = 0.63$\pm$0.02 for $n_{in}$ = 20.  This degradation may be partly explained by the fact that the stored and retrieved pulse is longer than the blockade time $\tau _b$, such that when $n_{in}$ increases, there is an increasing probability of having more than 1 photon in the retrieved pulse. Another  potential cause is the generation of pollutants in the cloud \cite{Bienias2020}. These pollutants can be spurious Rydberg excitations in $|r\rangle$ created by the reabsorbed scattered  photons, which cannot be retrieved in the detection mode and block the creation of retrievable polaritons. This type of pollutants is short-lived because they are in resonance with the control field. However, they could also decay to other Rydberg level not coupled to the control laser, which leads to long lived pollutants that can block the creation of Rydberg polaritons for several trials. To our knowledge, a proper theory describing the effect of pollutants on the indistinguishability and quality of single photons generated by Rydberg EIT with storage has not been proposed yet.

The exact nature and behavior of pollutants is still an open question, despite the existence of some models accounting for experimental observations \cite{Bienias2020, OrnelasHuerta2020}. One relevant open question is in regards to pollutants and optical potentials. Since most Rydberg atoms experience an anti-trap potential with the common dipole trap wavelengths used, we expect pollutants to experience a repulsive force each dipole trap cycle. The exact acceleration acquired depends heavily on $n$ and is roughly proportional to $~(n^{*})^7$ \cite{Lai2018}, were $n^*$ is the effective quantum number. Consequently, atoms capable of generating larger blockades are expelled from the interaction region faster, while lower lying Rydberg atoms block a smaller region, however can stay for longer in the interaction zone. For example, we expect that an atom with $n=50$ \cite{Zhang2011b} to be expelled from the trap in approximately 1 $\mu$s, while atoms with $n\lesssim 36$ being able to remain in the interaction zone. Those excitations have lifetimes of several $\mu$s and would be able to persist for several trials.

In summary, we observed that despite the fact that they exhibit a lower $g^{(2)}_{\Delta t}$,  the photons generated with the EIT protocol show lower  indistinguishability factors and HOM visibilities than those generated by the OR protocol. Moreover, the quality of the single photon degrades faster with $n_{in}$ . These effects may be due to the presence of scattering induced decoherence and generation of pollutants. However, more measurements and a full model of decoherence in Rydberg EIT storage would be needed to quantify the consequences of the two effects. 

\section{Conclusion}

We conducted an extensive study of indistinguishability for single photons generated by cold ensembles of Rydberg atoms. We performed HOM interferometry between single photons and weak coherent states and compared two excitation schemes, namely OR and EIT storage, providing the figures of merit for the single photon generation and characterization in each of those schemes. We concluded that the OR scheme generates single photons with a higher degree of indistinguishability than those generated through stored Rydberg EIT, presenting up to 89 $\%$ of indistinguishability for the full pulse, and up to $98\%$ of indistinguishability for a smaller detection window. We discussed the impact of extending the window of observation on the indistinguishability and concluded that, despite higher values of indistinguishability are achieved for smaller windows of observations, the single photons remain highly  indistinguishable for the whole duration of the pulse. These observations are specifically interesting for applications in quantum communications, where extending the window of observation implies a gain in efficiency and distant entanglement rate. 

\section*{acknowledgements} The authors would like to thank Emanuele Distante for interesting discussions. This project received funding from the  the Government of Spain (PID2019-106850RB-I00; Severo Ochoa CEX2019-000910-S), Fundació Cellex, Fundació Mir-Puig, from   Generalitat de Catalunya (CERCA, AGAUR), and from the Gordon and Betty Moore Foundation through Grant No. GBMF7446 to H. d. R.

%

\bibliographystyle{apsrev4-1}
\end{document}